\documentclass{ws-mpla}
\def\Journal#1#2#3#4{{#1} {\bf #2}, #3 (#4)}

\def\PU{\em Phys. Usp.}
\def\RNC{\em Rivista Nuovo Cimento}

\def\NIMA{{\em Nucl. Instrum. Methods} A}
\def\NPB{{\em Nucl. Phys.} B}
\def\PLB{{\em Phys. Lett.}  B}
\def\PRL{\em Phys. Rev. Lett.}
\def\PRD{{\em Phys. Rev.} D}

\def\GaC{\em Gravitation and Cosmology}
\def\GaCS{{\em Gravitation and Cosmology} Supplement}

\def\JETPL{\em JETP Lett.}
\def\PAN{\em Phys.Atom.Nucl.}
\def\CQG{\em Class. Quantum Grav.}
\def\APJ{\em Astrophys. J.}
\def\SCI{\em Science}
\def\MPLA{{\em Mod. Phys. Lett.}  A}
\def\IJTP{\em Int. J. Theor. Phys.}
\def\IJMPA{{\em Int. J. Mod. Phys.}  A}
\def\IJMPD{{\em Int. J. Mod. Phys.}  D}
\def\NJP{\em New J. of Phys.}
\def\ARAA{\em Ann. Rev. Astron. Astrophys.}
\def\AIPCP{\em AIP Conf. Proc.}
\def\JHEP{\em JHEP}
\def\JCAP{\em JCAP}
\def\EPHJ{\em Eur.Phys.J}
\def\JPCS{{\em J. Phys.:} Conf. Ser.}
\def\BWP{\em Bled Workshops in Physics}

\def\s{{\,\rm s}}
\def\g{{\,\rm g}}
\def\eV{\,{\rm eV}}
\def\keV{\,{\rm keV}}
\def\MeV{\,{\rm MeV}}
\def\GeV{\,{\rm GeV}}
\def\TeV{\,{\rm TeV}}
\def\sv{\left<\sigma v\right>}
\def\({\left(}
\def\){\right)}
\def\cm{{\,\rm cm}}
\def\K{{\,\rm K}}
\def\kpc{{\,\rm kpc}}
\def\beq{\begin{equation}}
\def\eeq{\end{equation}}
\def\bea{\begin{eqnarray}}
\def\eea{\end{eqnarray}}

\begin{document}

\markboth{M.Yu.KHLOPOV}
{PHYSICS OF DARK MATTER IN THE LIGHT OF DARK ATOMS}

\catchline{}{}{}{}{}

\title{PHYSICS OF DARK MATTER IN THE LIGHT OF DARK ATOMS
}

\author{MAXIM YU. KHLOPOV}

\address{National Research Nuclear University "Moscow Engineering Physics Institute" and \\
    Centre for Cosmoparticle Physics "Cosmion" 115409 Moscow, Russia \\
APC laboratory 10, rue Alice Domon et L\'eonie Duquet \\75205
Paris Cedex 13, France\\
khlopov@apc.univ-paris7.fr}

\maketitle

\pub{Received (Day Month Year)}{Revised (Day Month Year)}

\begin{abstract}
Direct searches for dark matter lead to serious problems for simple models with stable neutral Weakly Interacting Massive Particles (WIMPs) as candidates for dark matter. A possibility is discussed that new stable quarks and charged leptons exist and are hidden from detection, being bound in neutral dark atoms of composite dark matter. Stable -2 charged particles $O^{--}$ are bound with primordial helium in O-helium (OHe) atoms, being specific nuclear interacting form of composite Warmer than Cold dark matter. Slowed down in the terrestrial matter, OHe is elusive for
direct methods of underground dark matter detection based on the
search for effects of nuclear recoil in WIMP-nucleus collisions. The positive results of DAMA experiments can be explained as annual modulation of radiative capture of O-helium by nuclei. In the framework of this approach test of DAMA results in detectors with other chemical content becomes a nontrivial task, while the experimental search of stable charged particles at LHC or in cosmic rays acquires a meaning of direct test for composite dark matter scenario.

\keywords{Elementary particles; dark matter; early universe; nuclear reactions; radiative capture;
large-scale structure of universe.}
\end{abstract}

\ccode{PACS Nos.: include PACS Nos.}

\section{Introduction}
According to the modern cosmology, the dark matter, corresponding to
$\sim 25\%$ of the total cosmological density, is nonbaryonic and
consists of new stable particles. Such particles (see e.g.\cite{book,Cosmoarcheology,Bled06,Bled07,newBook}
for review and reference) should
be stable, saturate the measured dark matter density and decouple
from plasma and radiation at least before the beginning of matter
dominated stage. The easiest way to satisfy these conditions is to
involve neutral elementary weakly interacting particles. However it
is not the only particle physics solution for the dark matter
problem and more evolved models of self-interacting dark matter are
possible. In particular, new stable particles may possess new U(1)
gauge charges and bind by Coulomb-like forces in composite dark
matter species. Such dark atoms would look nonluminous, since they
radiate invisible light of U(1) photons. Historically mirror matter
(see\cite{book,Okun} for review and references) seems to be the
first example of such a nonluminous atomic dark matter.

Glashow's tera-helium\cite{Glashow} has offered a new solution for
dark atoms of dark matter. Tera-U-quarks with electric charge +2/3
formed stable (UUU) +2 charged "clusters" that formed with two -1
charged tera-electrons E neutral [(UUU)EE] tera-helium "atoms" that
behaved like Weakly Interacting Massive Particles (WIMPs). The main
problem for this solution was to suppress the abundance of
positively charged species bound with ordinary electrons, which
behave as anomalous isotopes of hydrogen or helium. This problem
turned to be unresolvable\cite{Fargion:2005xz}, since the model\cite{Glashow}
predicted stable tera-electrons $E^-$ with charge -1.
As soon as primordial helium is formed in the Standard Big Bang
Nucleosynthesis (SBBN) it captures all the free $E^-$ in positively
charged $(He E)^+$ ion, preventing any further suppression of
positively charged species. Therefore, in order to avoid anomalous
isotopes overproduction, stable particles with charge -1 (and
corresponding antiparticles) should be absent, so that stable
negatively charged particles should have charge -2 only.

Elementary particle frames for heavy stable -2 charged species are
provided by: (a) stable "antibaryons" $\bar U \bar U \bar U$ formed
by anti-$U$ quark of fourth generation\cite{Q,I,lom,KPS06,Belotsky:2008se,Khlopov:2006dk}
(b) AC-leptons\cite{Khlopov:2006dk,5,FKS,Khlopov:2006uv}, predicted in the
extension \cite{5} of standard model, based on the approach of
almost-commutative geometry\cite{bookAC}.  (c) Technileptons and
anti-technibaryons \cite{KK} in the framework of walking technicolor
models (WTC)\cite{Sannino:2004qp,Hong:2004td,Dietrich:2005jn,Dietrich:2005wk,Gudnason:2006ug,Gudnason:2006yj}. (d) Finally, stable charged
clusters $\bar u_5 \bar u_5 \bar u_5$ of (anti)quarks $\bar u_5$ of
5th family can follow from the approach, unifying spins and charges\cite{Norma,Norma2,Norma3,Norma4,Norma5}. Since all these models also predict corresponding +2
charge antiparticles, cosmological scenario should provide mechanism
of their suppression, what can naturally take place in the
asymmetric case, corresponding to excess of -2 charge species,
$O^{--}$. Then their positively charged antiparticles can
effectively annihilate in the early Universe.

If new stable species belong to non-trivial representations of
electroweak SU(2) group, sphaleron transitions at high temperatures
can provide the relationship between baryon asymmetry and excess of
-2 charge stable species, as it was demonstrated in the case of WTC
in\cite{KK,Levels1,KK2,unesco,iwara,I2}.

 After it is formed
in the Standard Big Bang Nucleosynthesis (SBBN), $^4He$ screens the excessive
$O^{--}$ charged particles in composite $(^4He^{++}O^{--})$ {\it
O-helium} ($OHe$) ``atoms''\cite{I}.

In all the considered forms of O-helium, $O^{--}$ behaves either as lepton or
as specific "heavy quark cluster" with strongly suppressed hadronic
interaction. Therefore O-helium interaction with matter is
determined by nuclear interaction of $He$. These neutral primordial
nuclear interacting species can play the role of a nontrivial form of strongly
interacting dark matter\cite{Starkman,Wolfram,Starkman2,Javorsek,Mitra,Mack,McGuire:2001qj,McGuire2,ZF}, giving rise to a Warmer than
Cold dark matter scenario\cite{Levels,Levels1,KK2}.

Here after a brief review of possible charged constituents of dark
atoms, we concentrate on the properties of OHe atoms, their interaction with matter and qualitative picture of OHe cosmological evolution\cite{I,Levels,FKS,KK,unesco,Khlopov:2008rp,KhlopovPHE} and observable effects. We show\cite{DMDA} that interaction of OHe with nuclei in
underground detectors can  explain positive results
of dark matter searches in DAMA/NaI (see for review\cite{Bernabei:2003za})
and DAMA/LIBRA\cite{Bernabei:2008yi}
experiments by annual modulations of radiative capture of O-helium, resolving the controversy
between these results and the results of other experimental groups.
\section{\label{asymmetry} Charged constituents of composite dark matter}

\subsection{Problem of tera-fermion composite dark matter}
Glashow's Tera-helium Universe was first inspiring example of the
composite dark matter scenario. $SU(3)_c \times SU(2) \times SU(2)'
\times U(1)$ gauge model\cite{Glashow} was aimed to explain the origin of the neutrino mass and to solve the problem of strong CP-violation in QCD. New extra $SU(2)'$ symmetry acts on three heavy generations of
tera-fermions  linked with the light fermions by $CP'$
transformation. $SU(2)'$ symmetry breaking at TeV scale makes
tera-fermions much heavier than their light partners. Tera-fermion
mass spectrum is the same as for light generations, but all the
masses are scaled by the same factor of about $10^6$. Thus the
masses of lightest heavy particles are in {\it tera}-eV (TeV) range,
explaining their name.

Glashow's model\cite{Glashow} takes into account
that
 very heavy quarks $Q$ (or antiquarks $\bar Q$) can form bound states with other heavy quarks
 (or antiquarks) due to their Coulomb-like QCD attraction, and the binding energy of these states
 substantially exceeds the binding energy of QCD confinement.
Then stable $(QQq)$ and $(QQQ)$ baryons can exist.

According to\cite{Glashow} primordial heavy quark $U$ and heavy
electron $E$ are stable and
may form a neutral $(UUUEE)$ "atom"
with $(UUU)$ hadron as nucleus and two $E^-$s as "electrons". The
gas of such "tera-helium atoms" was proposed in\cite{Glashow} as a candidate for a
WIMP-like dark matter.

The problem of such scenario is an
inevitable presence of "products of incomplete combustion" and the
necessity to decrease their abundance.

Unfortunately, as it was shown in\cite{Fargion:2005xz}, this
picture of Tera-helium Universe can not be realized.

When ordinary $^4$He is formed in Big Bang
Nucleosynthesis, it binds all the free
$E^-$ into positively charged $(^4HeE^-)^+$ "ions". This puts
Coulomb barrier for any successive $E^-E^+$ annihilation or any
effective $EU$ binding. It removes  a possibility to suppress the abundance of
unwanted tera-particle species (like $(eE^+)$, $(^4He Ee)$ etc).
For instance the remaining abundance of $(eE^+)$ and $(^4HeE^-e)$ exceeds the terrestrial upper limit for anomalous hydrogen by
{\it 27 orders} of magnitude\cite{Fargion:2005xz}.

\subsection{Composite dark matter from almost commutative geometry}
The AC-model is based on the specific mathematical approach of
unifying general relativity, quantum mechanics and gauge symmetry\cite{5,bookAC}.
This realization naturally embeds the Standard model, both
reproducing its gauge symmetry and Higgs mechanism with prediction of a Higgs boson mass. AC model
 is in some sense alternative to SUSY, GUT and superstring extension of Standard model. The AC-model\cite{5} extends the fermion content of the Standard
model by two heavy particles, $SU(2)$ electro-weak singlets, with opposite electromagnetic charges.
Each of them has its own antiparticle. Having no other gauge charges of Standard model,
these particles (AC-fermions) behave as heavy stable leptons with
charges $-2e$ and $+2e$, called $A^{--}$ and $C^{++}$, respectively.

Similar to the Tera-helium Universe, AC-lepton relics from
intermediate stages of a multi-step process towards a final $(AC)$
atom formation must survive in the present Universe. In spite of the assumed excess of
particles ($A^{--}$ and $C^{++}$) the abundance of relic
antiparticles ($\bar A^{++}$ and $\bar C^{--}$) is not negligible.
There may be also a significant fraction of $A^{--}$ and $C^{++}$, which remains
unbound after recombination process of these particles into $(AC)$ atoms took place. As soon as $^4He$ is formed in Big
Bang nucleosynthesis, the primordial component of free anion-like AC-leptons
($A^{--}$) is mostly trapped in the first three minutes into a
neutral O-helium atom $^4He^{++}A^{--}$.
O-helium is able to capture free $C^{++}$ creating $(AC)$ atoms and releasing $^4He$ back. In the same way the annihilation of antiparticles speeds up. $C^{++}$-O-helium reactions stop, when their timescale exceeds a cosmological time, leaving O-helium and $C^{++}$ relics in the Universe. The catalytic reaction of O-helium with $C^{++}$ in the dense matter bodies provides successive
$(AC)$ binding that suppresses terrestrial
anomalous isotope abundance below the experimental upper limit. Due to screened charge of AC-atoms they have WIMP-like interaction with the ordinary matter. Such WIMPs are inevitably accompanied by a tiny component of nuclear interacting O-helium.

\subsection{Stable charged techniparticles in Walking Technicolor}

The minimal walking technicolor model\cite{Sannino:2004qp,Hong:2004td,Dietrich:2005jn,Dietrich:2005wk,Gudnason:2006ug,Gudnason:2006yj}
has two techniquarks, i.e. up $U$ and down $D$, that transform
under the adjoint representation of an $SU(2)$ technicolor gauge
group. The six
Goldstone bosons $UU$, $UD$, $DD$ and their corresponding
antiparticles carry technibaryon number since they are made of
two techniquarks or two anti-techniquarks. This means that if there is no
processes violating the technibaryon number the lightest
technibaryon will be stable.

The electric charges of $UU$, $UD$,
and $DD$ are given in general by $q+1$, $q$, and $q-1$
respectively, where $q$ is an arbitrary real number. The model requires in addition
the existence of a fourth family of leptons, i.e. a ``new
neutrino'' $\nu'$ and a ``new electron'' $\zeta$. Their electric charges are in
terms of $q$ respectively $(1-3q)/2$ and $(-1-3q)/2$.

There are three possibilities for a scenario of dark atoms of dark matter. The first one is to have an excess of $\bar{U}\bar{U}$ (charge $-2$).
The technibaryon
number $TB$ is conserved and therefore $UU$ (or $\bar{U}\bar{U}$) is
stable. The second possibility is to
have excess of $\zeta$ that also has $-2$ charge and is
stable, if $\zeta$ is lighter than $\nu'$ and technilepton number $L'$  is conserved. In the both cases
stable particles with $-2$ electric charge have substantial relic
densities and can capture $^4He^{++}$ nuclei to form a neutral techni-O-helium
atom.
Finally there is a
possibility to have both  $L'$ and $TB$ conserved. In this case, the dark matter would be composed
of bound atoms $(^4He^{++}\zeta^{--})$ and $(\zeta^{--}(U U )^{++})$. In the latter case the excess of $\zeta^{--}$ should be larger, than the excess of $(U U )^{++})$, so that WIMP-like $(\zeta^{--}(U U )^{++})$ is subdominant at the dominance of nuclear interacting techni-O-helium.

The technicolor and the
Standard Model particles are in thermal equilibrium as long as the
timescale of the weak (and color) interactions is smaller than the
cosmological time. The sphalerons allow violation of  $TB$, of baryon number $B$, of lepton number $L$ and  $L'$ as
long as the temperature of the Universe exceeds the electroweak scale.
It was shown in\cite{KK} that there is a balance between the excess of techni(anti)baryons, $(\bar{U}\bar{U})^{--}$,
technileptons $\zeta^{--}$ or of the both over the corresponding particles ($UU$ and/or $\zeta^{++}$) and the observed baryon asymmetry
of the Universe. It was also shown the there are parameters of the model, at which this asymmetry has
proper sign and value, explaining the dark matter density.

\subsection{\label{4generation} Stable particles of 4th generation matter}
Modern precision data
on the parameters of the Standard model do not exclude\cite{Maltoni:1999ta}
the existence of
the  4th generation of quarks and leptons. The 4th generation follows from heterotic string phenomenology and
its difference from the three known light generations can be
explained by a new conserved charge, possessed only by
its quarks and leptons\cite{Q,I,Belotsky:2000ra,Belotsky:2005uj,Belotsky:2004st}. Strict conservation of this charge makes the
lightest particle of 4th family (neutrino) absolutely
stable, but it was shown in\cite{Belotsky:2000ra,Belotsky:2005uj,Belotsky:2004st} that this neutrino cannot be the dominant form of the dark matter.
The same conservation law requires the lightest quark to be long living
\cite{Q,I}. In principle the lifetime of $U$ can exceed the age of the
Universe, if $m_U<m_D$\cite{Q,I}.
Provided that sphaleron transitions establish excess of $\bar U$ antiquarks at the observed baryon asymmetry
 $(\bar U \bar U \bar U)$ can be formed and bound with $^4He$ in atom-like state
of O-helium\cite{Khlopov:2006dk}.

In the successive discussion of OHe dark matter we generally don't specify the type of $-2$ charged particle, denoting it as $O^{--}$.

\section{OHe atoms and their interaction with nuclei}
The structure of OHe atom follows from the general
analysis of the bound states of $O^{--}$ with nuclei.

Consider a simple model\cite{Cahn,Pospelov,Kohri}, in which the nucleus is
regarded as a sphere with uniform charge density and in which the
mass of the $O^{--}$ is assumed to be much larger than that of the
nucleus. Spin dependence is also not taken into account so that both
the particle and nucleus are considered as scalars. Then the
Hamiltonian is given by
\begin{equation}
    H=\frac{p^2}{2 A m_p} - \frac{Z Z_x \alpha}{2 R} + \frac{Z Z_x \alpha}{2 R} \cdot (\frac{r}{R})^2,
\end{equation}
for short distances $r<R$ and
\begin{equation}
    H=\frac{p^2}{2 A m_p} - \frac{Z Z_x \alpha}{R},
\end{equation}
for long distances $r>R$, where $\alpha$ is the fine structure
constant, $R = d_o A^{1/3} \sim 1.2 A^{1/3} /(200 MeV)$ is the
nuclear radius, $Z$ is the electric charge of nucleus and $Z_x=2$ is
the electric charge of negatively charged particle $X^{--}$. Since
$A m_p \ll M_X$ the reduced mass is $1/m= 1/(A m_p) + 1/M_X \approx
1/(A m_p)$.

For small nuclei the Coulomb binding energy is like in hydrogen atom
and is given by
\begin{equation}
    E_b=\frac{1}{2} Z^2 Z_x^2 \alpha^2 A m_p.
\end{equation}

For large nuclei $X^{--}$ is inside nuclear radius and the harmonic
oscillator approximation is valid for the estimation of the binding
energy
\begin{equation}
    E_b=\frac{3}{2}(\frac{Z Z_x \alpha}{R}-\frac{1}{R}(\frac{Z Z_x \alpha}{A m_p R})^{1/2}).
\label{potosc}
\end{equation}

For the intermediate regions between these two cases with the use of
trial function of the form $\psi \sim e^{- \gamma r/R}$ variational
treatment of the problem\cite{Cahn,Pospelov,Kohri} gives
\begin{equation}
    E_b=\frac{1}{A m_p R^2} F(Z Z_x \alpha A m_p R ),
\end{equation}
where the function $F(a)$ has limits
\begin{equation}
    F(a \rightarrow 0) \rightarrow \frac{1}{2}a^2  - \frac{2}{5} a^4
\end{equation}
and
\begin{equation}
    F(a \rightarrow \infty) \rightarrow \frac{3}{2}a  - (3a)^{1/2},
\end{equation}
where $a = Z Z_x \alpha A m_p R$. For $0 < a < 1$ the Coulomb model
gives a good approximation, while at $2 < a < \infty$ the harmonic
oscillator approximation is appropriate.

In the case of OHe $a = Z Z_x \alpha A m_p R \le 1$, what proves its
Bohr-atom-like structure, assumed in\cite{I,lom,Khlopov:2006dk,KK,unesco,iwara,I2}.
The radius of Bohr orbit in these ``atoms"
\cite{I,Levels} $r_{o} \sim 1/(Z_{o} Z_{He}\alpha m_{He}) \approx 2
\cdot 10^{-13} \cm $.
However, the size of
He nucleus, rotating around $O^{--}$ in this Bohr atom, turns out to be of
the order and even a bit larger than the radius $r_o$ of its Bohr
orbit, and the corresponding correction to the binding energy due to
non-point-like charge distribution in He is significant.

Bohr atom like structure of OHe seems to provide a possibility to
use the results of atomic physics for description of OHe interaction
with matter. However, the situation is much more complicated. OHe
atom is similar to the hydrogen, in which electron is hundreds times
heavier, than proton, so that it is proton shell that surrounds
"electron nucleus". Nuclei that interact with such "hydrogen" would
interact first with strongly interacting "protonic" shell and such
interaction can hardly be treated in the framework of perturbation
theory. Moreover in the description of OHe interaction the account
for the finite size of He, which is even larger than the radius of
Bohr orbit, is important. One should consider, therefore, the
analysis, presented below, as only a first step approaching true
nuclear physics of OHe.

The approach of\cite{Levels,Levels1} assumes the following
picture of OHe interaction with nuclei: OHe is a neutral atom in the ground state,
perturbed  by Coulomb and nuclear forces of the approaching nucleus.
The sign of OHe polarization changes with the distance: at larger distances Stark-like effect takes place - nuclear Coulomb force polarizes OHe so that  nucleus is attracted by the induced dipole moment of OHe, while as soon as the perturbation by nuclear force starts to dominate the nucleus polarizes OHe in the opposite way so that He is situated more close to the nucleus, resulting in the repulsive effect of the helium shell of OHe.
When helium is completely merged with the nucleus the interaction is
reduced to the oscillatory potential of $O^{--}$ with
homogeneously charged merged nucleus with the charge $Z+2$.

\begin{figure}[ph]
\centerline{\psfig{file=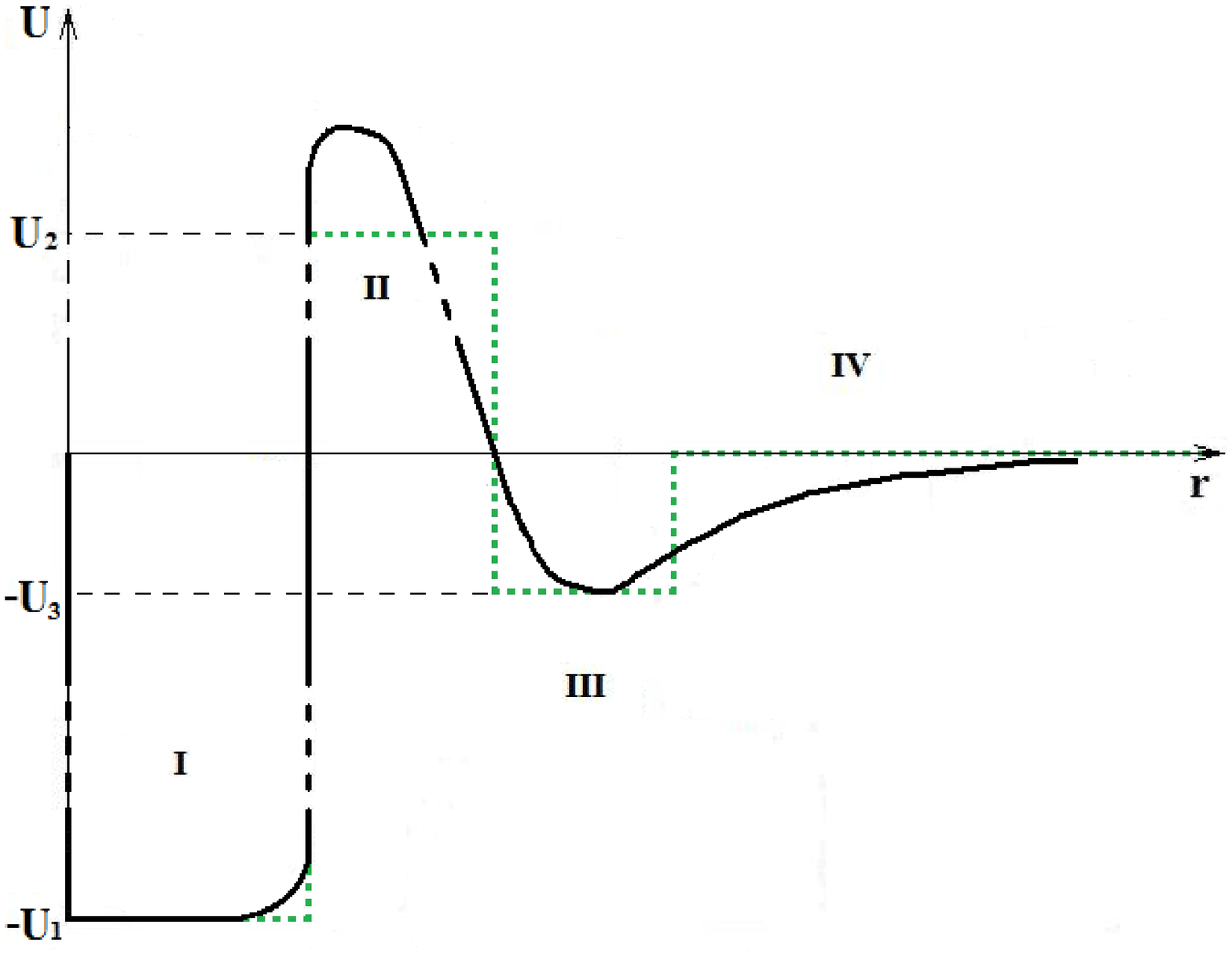,width=4.0in}}
\vspace*{8pt}
\caption{The potential of OHe-nucleus system and its rectangular well approximation.\protect\label{pic1}}
\end{figure}

Therefore OHe-nucleus potential has qualitative feature, presented on Fig.~\ref{pic1}:
the potential well $U_3$ at large distances (regions III-IV) is changed by a potential wall $U_2$ in region II. The existence of this potential barrier causes suppression of reactions with transition of OHe-nucleus system to levels in the potential well $U_1$ of the region I. It results in the dominance of elastic scattering while transitions to levels in the shallow well (regions III-IV) should dominate in reactions of OHe-nucleus capture.

On the other hand, O-helium, being an $\alpha$-particle with screened electric charge,
can catalyze nuclear transformations, which can influence primordial
light element abundance and cause primordial heavy element
formation. It is especially important for quantitative estimation of
role of OHe in Big Bang Nucleosynthesis and in stellar evolution.
These effects need a special detailed and complicated
study and
this work is under way. Our first steps
in the approach to OHe nuclear physics seem to support the qualitative
picture of OHe cosmological evolution described in\cite{I,Levels,FKS,KK,Levels1,unesco,Khlopov:2008rp}
and based on the dominant role of elastic
collisions in OHe interaction with baryonic matter.

\section{Some features of O-helium Universe}

\subsection{Large Scale structure formation by OHe dark matter}
Due to elastic nuclear interactions of its helium constituent with nuclei in
the cosmic plasma, the O-helium gas is in thermal equilibrium with
plasma and radiation on the Radiation Dominance (RD) stage, while
the energy and momentum transfer from plasma is effective. The
radiation pressure acting on the plasma is then transferred to
density fluctuations of the O-helium gas and transforms them in
acoustic waves at scales up to the size of the horizon.

At temperature $T < T_{od} \approx 1 S_3^{2/3}\eV$ the energy and
momentum transfer from baryons to O-helium is not effective
\cite{I,KK} because $$n_B \sv (m_p/m_o) t < 1,$$ where $m_o$ is the
mass of the $OHe$ atom and $S_3= m_o/(1 \TeV)$. Here \beq \sigma
\approx \sigma_{o} \sim \pi r_{o}^2 \approx
10^{-25}\cm^2\label{sigOHe}, \eeq and $v = \sqrt{2T/m_p}$ is the
baryon thermal velocity. Then O-helium gas decouples from plasma. It
starts to dominate in the Universe after $t \sim 10^{12}\s$  at $T
\le T_{RM} \approx 1 \eV$ and O-helium ``atoms" play the main
dynamical role in the development of gravitational instability,
triggering the large scale structure formation. The composite nature
of O-helium determines the specifics of the corresponding dark
matter scenario.

At $T > T_{RM}$ the total mass of the $OHe$ gas with density $\rho_d
= (T_{RM}/T) \rho_{tot} $ is equal to
$$M=\frac{4 \pi}{3} \rho_d t^3 = \frac{4 \pi}{3} \frac{T_{RM}}{T} m_{Pl}
(\frac{m_{Pl}}{T})^2$$ within the cosmological horizon $l_h=t$. In
the period of decoupling $T = T_{od}$, this mass  depends strongly
on the O-helium mass $S_3$ and is given by \cite{KK}\beq M_{od} =
\frac{T_{RM}}{T_{od}} m_{Pl} (\frac{m_{Pl}}{T_{od}})^2 \approx 2
\cdot 10^{44} S^{-2}_3 \g = 10^{11} S^{-2}_3 M_{\odot}, \label{MEPm}
\eeq where $M_{\odot}$ is the solar mass. O-helium is formed only at
$T_{o}$ and its total mass within the cosmological horizon in the
period of its creation is $M_{o}=M_{od}(T_{od}/T_{o})^3 = 10^{37}
\g$.

On the RD stage before decoupling, the Jeans length $\lambda_J$ of
the $OHe$ gas was restricted from below by the propagation of sound
waves in plasma with a relativistic equation of state
$p=\epsilon/3$, being of the order of the cosmological horizon and
equal to $\lambda_J = l_h/\sqrt{3} = t/\sqrt{3}.$ After decoupling
at $T = T_{od}$, it falls down to $\lambda_J \sim v_o t,$ where $v_o
= \sqrt{2T_{od}/m_o}.$ Though after decoupling the Jeans mass in the
$OHe$ gas correspondingly falls down
$$M_J \sim v_o^3 M_{od}\sim 3 \cdot 10^{-14}M_{od},$$ one should
expect a strong suppression of fluctuations on scales $M<M_o$, as
well as adiabatic damping of sound waves in the RD plasma for scales
$M_o<M<M_{od}$. It can provide some suppression of small scale
structure in the considered model for all reasonable masses of
O-helium. The significance of this suppression and its effect on the
structure formation needs a special study in detailed numerical
simulations. In any case, it can not be as strong as the free
streaming suppression in ordinary Warm Dark Matter (WDM) scenarios,
but one can expect that qualitatively we deal with Warmer Than Cold
Dark Matter model.

At temperature $T < T_{od} \approx 1 S_3^{2/3} \keV$ the energy and
momentum transfer from baryons to O-helium is not effective\cite{I,Levels,Levels1}
and O-helium gas decouples from plasma. It
starts to dominate in the Universe after $t \sim 10^{12}\s$  at $T
\le T_{RM} \approx 1 \eV$ and O-helium ``atoms" play the main
dynamical role in the development of gravitational instability,
triggering the large scale structure formation. The composite nature
of O-helium determines the specifics of the corresponding warmer than cold dark
matter scenario.

Being decoupled from baryonic matter, the $OHe$ gas does not follow
the formation of baryonic astrophysical objects (stars, planets,
molecular clouds...) and forms dark matter halos of galaxies. It can
be easily seen that O-helium gas is collisionless for its number
density, saturating galactic dark matter. Taking the average density
of baryonic matter one can also find that the Galaxy as a whole is
transparent for O-helium in spite of its nuclear interaction. Only
individual baryonic objects like stars and planets are opaque for
it.

\subsection{Anomalous component of cosmic rays}
O-helium atoms can be destroyed in astrophysical processes, giving
rise to acceleration of free $O^{--}$ in the Galaxy.

O-helium can be ionized due to nuclear interaction with cosmic rays\cite{I,I2}.
Estimations\cite{I,Mayorov} show that for the number
density of cosmic rays $ n_{CR}=10^{-9}\cm^{-3}$ during the age of
Galaxy a fraction of about $10^{-6}$ of total amount of OHe is
disrupted irreversibly, since the inverse effect of recombination of
free $O^{--}$ is negligible. Near the Solar system it leads to
concentration of free $O^{--}$ $ n_{O}= 3 \cdot 10^{-10}S_3^{-1}
\cm^{-3}.$ After OHe destruction free $O^{--}$ have momentum of
order $p_{O} \cong \sqrt{2 \cdot m_{o} \cdot I_{o}} \cong 2 \GeV
S_3^{1/2}$ and velocity $v/c \cong 2 \cdot 10^{-3} S_3^{-1/2}$ and
due to effect of Solar modulation these particles initially can
hardly reach Earth\cite{KK2,Mayorov}. Their acceleration by Fermi
mechanism or by the collective acceleration forms power spectrum of
$O^{--}$ component at the level of $O/p \sim n_{O}/n_g = 3 \cdot
10^{-10}S_3^{-1},$ where $n_g \sim 1 \cm^{-3}$ is the density of
baryonic matter gas.

At the stage of red supergiant stars have the size $\sim 10^{15}
\cm$ and during the period of this stage$\sim 3 \cdot 10^{15} \s$,
up to $\sim 10^{-9}S_3^{-1}$ of O-helium atoms per nucleon can be
captured\cite{KK2,Mayorov}. In the Supernova explosion these OHe
atoms are disrupted in collisions with particles in the front of
shock wave and acceleration of free $O^{--}$ by regular mechanism
gives the corresponding fraction in cosmic rays. However, this
picture needs detailed analysis, based on the development of OHe
nuclear physics and numerical studies of OHe evolution in the
stellar matter.

If these mechanisms of $O^{--}$ acceleration are effective, the
anomalous low $Z/A$ component of $-2$ charged $O^{--}$ can be
present in cosmic rays at the level $O/p \sim n_{O}/n_g \sim
10^{-9}S_3^{-1},$ and be within the reach for PAMELA and AMS02
cosmic ray experiments.

In the framework of Walking Technicolor model the excess of both
stable $\zeta^{--}$ and $(UU)^{++}$ is possible\cite{KK2}, the latter
being two-three orders of magnitude smaller, than the former. It
leads to the two-component composite dark matter scenario with the
dominant OHe accompanied by a subdominant WIMP-like component of
$(\zeta^{--}(U U )^{++})$ bound systems. Technibaryons and technileptons can
be metastable and decays of $\zeta^{--}$ and $(UU)^{++}$ can provide
explanation for anomalies, observed in high energy cosmic positron
spectrum by PAMELA and in high energy electron spectrum by FERMI and
ATIC.

\subsection{Positron annihilation and gamma lines in galactic
bulge}
Inelastic interaction of O-helium with the matter in the
interstellar space and its de-excitation can give rise to radiation
in the range from few keV to few  MeV. In the galactic bulge with
radius $r_b \sim 1 \kpc$ the number density of O-helium can reach
the value $n_o\approx 3 \cdot 10^{-3}/S_3 \cm^{-3}$ and the
collision rate of O-helium in this central region was estimated in
\cite{I2}: $dN/dt=n_o^2 \sigma v_h 4 \pi r_b^3 /3 \approx 3 \cdot
10^{42}S_3^{-2} \s^{-1}$. At the velocity of $v_h \sim 3 \cdot 10^7
\cm/\s$ energy transfer in such collisions is $\Delta E \sim 1 \MeV
S_3$. These collisions can lead to excitation of O-helium. If 2S
level is excited, pair production dominates over two-photon channel
in the de-excitation by $E0$ transition and positron production with
the rate $3 \cdot 10^{42}S_3^{-2} \s^{-1}$ is not accompanied by
strong gamma signal. According to\cite{Finkbeiner:2007kk} this rate
of positron production for $S_3 \sim 1$ is sufficient to explain the
excess in positron annihilation line from bulge, measured by
INTEGRAL (see\cite{integral} for review and references). If $OHe$
levels with nonzero orbital momentum are excited, gamma lines should
be observed from transitions ($ n>m$) $E_{nm}= 1.598 \MeV (1/m^2
-1/n^2)$ (or from the similar transitions corresponding to the case
$I_o = 1.287 \MeV $) at the level $3 \cdot 10^{-4}S_3^{-2}(\cm^2 \s
\MeV ster)^{-1}$.

It should be noted that the nuclear cross section of the O-helium
interaction with matter escapes the severe constraints\cite{McGuire:2001qj,McGuire2,ZF}
on strongly interacting dark matter particles
(SIMPs)\cite{Starkman,Wolfram,Starkman2,Javorsek,Mitra,Mack,McGuire:2001qj,McGuire2,ZF} imposed by the XQC experiment\cite{XQC,XQC1}. Therefore, a special strategy of direct O-helium  search
is needed, as it was proposed in\cite{Belotsky:2006fa}.

\subsection{O-helium in the terrestrial matter} The evident
consequence of the O-helium dark matter is its inevitable presence
in the terrestrial matter, which appears opaque to O-helium and
stores all its in-falling flux.

After they fall down terrestrial surface, the in-falling $OHe$
particles are effectively slowed down due to elastic collisions with
matter. Then they drift, sinking down towards the center of the
Earth with velocity \beq V = \frac{g}{n \sigma v} \approx 80 S_3
A_{med}^{1/2} \cm/\s. \label{dif}\eeq Here $A_{med} \sim 30$ is the average
atomic weight in terrestrial surface matter, $n=2.4 \cdot 10^{24}/A$
is the number of terrestrial atomic nuclei, $\sigma v$ is the rate
of nuclear collisions and $g=980~ \cm/\s^2$.

Near the Earth's surface, the O-helium abundance is determined by
the equilibrium between the in-falling and down-drifting fluxes.

At a depth $L$ below the Earth's surface, the drift timescale is
$t_{dr} \sim L/V$, where $V \sim 400 S_3 \cm/\s$ is the drift velocity and $m_o=S_3 \TeV$ is the mass of O-helium. It means that the change of the incoming flux,
caused by the motion of the Earth along its orbit, should lead at
the depth $L \sim 10^5 \cm$ to the corresponding change in the
equilibrium underground concentration of $OHe$ on the timescale
$t_{dr} \approx 2.5 \cdot 10^2 S_3^{-1}\s$.

The equilibrium concentration, which is established in the matter of
underground detectors at this timescale, is given by
\begin{equation}
    n_{oE}=n_{oE}^{(1)}+n_{oE}^{(2)}\cdot sin(\omega (t-t_0))
    \label{noE}
\end{equation}
with $\omega = 2\pi/T$, $T=1yr$ and
$t_0$ the phase.
So, there is a averaged concentration given by
\begin{equation}
    n_{oE}^{(1)}=\frac{n_o}{320S_3 A_{med}^{1/2}} V_{h}
\end{equation}
and the annual modulation of concentration characterized by the amplitude
\begin{equation}
    n_{oE}^{(2)}= \frac{n_o}{640S_3 A_{med}^{1/2}} V_E.
\end{equation}
Here $V_{h}$-speed of Solar System (220 km/s), $V_{E}$-speed of
Earth (29.5 km/s) and $n_{0}=3 \cdot 10^{-4} S_3^{-1} \cm^{-3}$ is the
local density of O-helium dark matter.

\section{OHe in the underground detectors}

The explanation\cite{Levels,DMDA} of the results of
DAMA/NaI\cite{Bernabei:2003za} and DAMA/LIBRA\cite{Bernabei:2008yi}
experiments is based on the idea that OHe,
slowed down in the matter of detector, can form a few keV bound
state with nucleus, in which OHe is situated \textbf{beyond} the
nucleus. Therefore the positive result of these experiments is
explained by annual modulation in reaction of radiative capture of OHe
\begin{equation}
A+(^4He^{++}O^{--}) \rightarrow [A(^4He^{++}O^{--})]+\gamma
\label{HeEAZ}
\end{equation}
by nuclei in DAMA detector.

To simplify the solution of Schrodinger equation the
potential was approximated in\cite{Levels,Levels1} by a rectangular potential, presented on Fig.~\ref{pic1}.
Solution of Schrodinger equation determines the condition, under
which a low-energy  OHe-nucleus bound state appears in the shallow well of the region
III and the range of nuclear parameters was found, at which OHe-sodium binding energy is in the interval 2-4 keV.


The rate of radiative capture of OHe by nuclei can be calculated\cite{Levels,DMDA}
with the use of the analogy with the radiative
capture of neutron by proton with the account for: i) absence of M1
transition that follows from conservation of orbital momentum and
ii) suppression of E1 transition in the case of OHe. Since OHe is
isoscalar, isovector E1 transition can take place in OHe-nucleus
system only due to effect of isospin nonconservation, which can be
measured by the factor $f = (m_n-m_p)/m_N \approx 1.4 \cdot
10^{-3}$, corresponding to the difference of mass of neutron,$m_n$,
and proton,$m_p$, relative to the mass of nucleon, $m_N$. In the
result the rate of OHe radiative capture by nucleus with atomic
number $A$ and charge $Z$ to the energy level $E$ in the medium with
temperature $T$ is given by
\begin{equation}
    \sigma v=\frac{f \pi \alpha}{m_p^2} \frac{3}{\sqrt{2}} (\frac{Z}{A})^2 \frac{T}{\sqrt{Am_pE}}.
    \label{radcap}
\end{equation}

Formation of OHe-nucleus bound system leads to energy release of its
binding energy, detected as ionization signal.  In the context of
our approach the existence of annual modulations of this signal in
the range 2-6 keV and absence of such effect at energies above 6 keV
means that binding energy $E_{Na}$ of Na-OHe system in DAMA experiment should
not exceed 6 keV, being in the range 2-4 keV. The amplitude of
annual modulation of ionization signal can reproduce the result of DAMA/NaI and DAMA/LIBRA
experiments for $E_{Na} = 3 \keV$. The
account for energy resolution in DAMA experiments\cite{DAMAlibra}
can explain the observed energy distribution of the signal from
monochromatic photon (with $E_{Na} = 3 \keV$) emitted in OHe
radiative capture.

At the corresponding nuclear parameters there is no binding
of OHe with iodine and thallium\cite{Levels}.

It should be noted that the results of DAMA experiment exhibit also
absence of annual modulations at the energy of MeV-tens MeV. Energy
release in this range should take place, if OHe-nucleus system comes
to the deep level inside the nucleus. This transition implies
tunneling through dipole Coulomb barrier and is suppressed below the
experimental limits.

For the chosen range of nuclear parameters, reproducing the results
of DAMA/NaI and DAMA/LIBRA, the results\cite{Levels} indicate that
there are no levels in the OHe-nucleus systems for heavy nuclei. In
particular, there are no such levels in Xe, what
seem to prevent direct comparison with DAMA results in
XENON100 experiment\cite{xenon}. The existence of such level in Ge and the comparison with the results of
CDMS\cite{Akerib:2005kh,Ahmed:2008eu,cdms} and CoGeNT\cite{cogent} experiments need special study. According to\cite{Levels} OHe should bind with O and Ca, what is of interest for interpretation of the signal, observed in CRESST-II experiment\cite{cresst}.

In the thermal equilibrium OHe capture rate is proportional to the temperature. Therefore it looks
like it is suppressed in cryogenic detectors by a factor of order
$10^{-4}$. However, for the size of cryogenic devices  less, than
few tens meters, OHe gas in them has the thermal velocity of the
surrounding matter and this velocity dominates in the relative velocity of OHe-nucleus system.
It gives the suppression relative to room temperature
only $\sim m_A/m_o$. Then the rate of OHe radiative capture in
cryogenic detectors is given by Eq.(\ref{radcap}), in which room
temperature $T$ is multiplied by factor $m_A/m_o$. Note that in the case of $T=70\K$ in CoGeNT experiment
relative velocity is determined by the thermal velocity of germanium nuclei, what leads to enhancement relative to cryogenic germanium detectors.
\section{Conclusions}
The existence of heavy stable particles is one of the popular solutions for the dark matter problem.
Usually they are considered to be electrically neutral. But potentially dark matter can be formed by
stable heavy charged particles bound in neutral atom-like states by Coulomb attraction.
Analysis of the cosmological data and atomic composition of the Universe gives the constrains
on the particle charge showing that  only $-2$
charged constituents, being trapped by primordial helium
in neutral O-helium states, can avoid the problem of overproduction of the anomalous isotopes of chemical elements, which are severely constrained by observations. Cosmological model of O-helium dark matter
can even explain puzzles of direct dark matter searches.

The proposed explanation is based on the mechanism of low energy
binding of OHe with nuclei. Within the uncertainty of nuclear
physics parameters there exists a range at which OHe binding energy
with sodium is in the interval 2-4 keV. Annual modulation in radiative capture of OHe to
this bound state leads to the corresponding energy release observed
as an ionization signal in DAMA/NaI and
DAMA/LIBRA experiments.


With the account for high sensitivity of the numerical results to
the values of nuclear parameters and for the approximations, made in
the calculations, the presented results can be considered only as an
illustration of the possibility to explain puzzles of dark matter
search in the framework of composite dark matter scenario. An
interesting feature of this explanation is a conclusion that the
ionization signal may
be absent in detectors containing light (e.g. $^3He$) or heavy (e.g. Xe) elements.
Therefore test of results of DAMA/NaI and
DAMA/LIBRA experiments by other experimental groups can become a
very nontrivial task. Recent indications to positive result in the matter of CRESST detector\cite{cresst},
in which OHe binding is expected together with absence of signal in xenon detector\cite{xenon}, may qualitatively favor the presented approach. For the same chemical content
an order of magnitude suppression in cryogenic detectors can explain why indications to positive effect in
CoGeNT experiment\cite{cogent} can be compatible with the constraints of CDMS experiment.

An inevitable consequence of the proposed explanation is appearance
in the matter of underground detectors anomalous
superheavy isotopes, having the mass roughly by $m_o$
larger, than ordinary isotopes of the corresponding elements.

It is interesting to note that in the framework of the presented approach
positive result of experimental search for WIMPs by effect of their
nuclear recoil would be a signature for a multicomponent nature of
dark matter. Such OHe+WIMPs multicomponent dark matter scenarios
naturally follow from AC model \cite{FKS} and can be realized in
models of Walking technicolor \cite{KK2}.

Stable $-2$ charge states ($O^{--}$) can be elementary like AC-leptons or technileptons,
or look like technibaryons. The latter, composed of techniquarks, reveal their structure at much higher energy scale and should be produced at LHC as
elementary species. The signature  for AC leptons and techniparticles is unique and distinctive what  allows
to separate them  from other hypothetical exotic particles.

Since simultaneous production of three $U \bar U$ pairs and
their conversion in two doubly charged quark clusters $UUU$
is suppressed, the only possibility to test the
models of composite dark matter from 4th generation in the collider experiments is a search for production of stable hadrons containing single $U$ or $\bar U$ like $Uud$ and $\bar U u$/$\bar U d$.

The presented approach sheds new light on the physical nature of
dark matter. Specific properties of dark atoms and their
constituents contain distinct features, by which they can be distinguished from other recent approaches to this problem\cite{Edward,Foot,Feng1,Feng2,Drob,Feldstein:2009tr,Bai,Feldstein:2009np,Fitzpatrick:2010em,Andreas:2010dz,Alves:2010dd,Barger:2010yn,Savage:2010tg,Hooper:2010uy,Chang:2010pr,Chang:2010en,Barger:2010gv,Fitzpatrick:2010br,Banks:2010eh,Feldstein:2010su,Gelmini,Aprile:2009zzd,Feng:2010gw,Dai,Jia}, and are challenging for the experimental search. The
development of quantitative description of OHe interaction with
matter confronted with the experimental data will provide the
complete test of the composite dark matter model. It challenges search for stable double charged particles at accelerators and cosmic rays as direct experimental probe for charged constituents of dark atoms of dark matter.

\section*{Acknowledgments}
I express my gratitude to K.M. Belotsky, D. Fargion, C. Kouvaris, A.G. Mayorov, E. Yu. Soldatov and C. Stephan for collaboration in obtaining the original results and to J.R. Cudell and A.S. Romaniouk for discussions.


\begin{thebibliography}{0}

\bibitem{book}  M.Yu. Khlopov
{\em Cosmoparticle physics} (World Scientific, Singapore, 1999).

\bibitem{Cosmoarcheology}  M.Yu. Khlopov in {\em Cosmion-94}, Eds. M.Yu.Khlopov et al.
(Editions frontieres, 1996) P. 67.
\bibitem{Bled06}
  M.~Y.~Khlopov, \Journal{\BWP}{7}{51}{2006}.

\bibitem{Bled07}
  M.~Y.~Khlopov, \Journal{\BWP}{8}{114}{2007}.

\bibitem{newBook}
 M.Yu. Khlopov
{\em Fundamentals of Cosmoparticle physics} (CISP-Springer, Cambridge, 2011).

\bibitem{Okun}
  L.~B.~Okun,\Journal{\PU}{50}{380}{2007}.

\bibitem{Glashow} S.~L.~Glashow,
  arXiv:hep-ph/0504287.

\bibitem{Fargion:2005xz}
  D. Fargion and M. Khlopov,
  arXiv:hep-ph/0507087.

\bibitem{Q}  
K.M.Belotsky {\it et al}, \Journal{\GaC}{11}{3}{2005}

\bibitem{I} M.Yu. Khlopov, \Journal{\JETPL}{83}{1}{2006}.
\bibitem{Levels}
  M.~Y.~Khlopov, A.~G.~Mayorov and E.~Y.~Soldatov,
  \Journal{\JPCS}{309}{012013}{2011}.

\bibitem{lom}
  K.~Belotsky {\it et al},
  arXiv:astro-ph/0602261.
\bibitem{KPS06}
  K.~Belotsky {\it et al}, \Journal{\GaC}{12}{1}{2006}.
\bibitem{Belotsky:2008se}
  K.~Belotsky {\it et al}, in {\it The Physics of Quarks: New Research.( Horizons in World Physics, V.265)} Eds. N. L. Watson and T. M. Grant, (NOVA Publishers, Hauppauge NY, 2009), p.19.

\bibitem{Khlopov:2006dk}
  M.~Y.~Khlopov,
  arXiv:astro-ph/0607048.












\bibitem{5} C.~A.~Stephan,
  arXiv:hep-th/0509213.

\bibitem{FKS} D.~Fargion {\it et al},
\Journal{\CQG}{23}{7305}{2006}.

\bibitem{Khlopov:2006uv}
  M.~Y.~Khlopov and C.~A.~Stephan,
  arXiv:astro-ph/0603187.

\bibitem{bookAC} A. Connes {\em Noncommutative Geometry} (Academic Press, London and San
Diego, 1994).




\bibitem{KK}
  M.~Y.~Khlopov and C.~Kouvaris,  \Journal{\PRD}{77}{065002}{2008}.

\bibitem{Sannino:2004qp}
F.~Sannino and K.~Tuominen,  \Journal{\PRD}{71}{051901}{2005}.
\bibitem{Hong:2004td}
  D.~K.~Hong {\it et al}, \Journal{\PLB}{597}{89}{2004}.
\bibitem{Dietrich:2005jn}
  D.~D.~Dietrich {\it et al}, \Journal{\PRD}{72}{055001}{2005}.
\bibitem{Dietrich:2005wk}
  D.~D.~Dietrich {\it et al}, \Journal{\PRD}{73}{037701}{2006}.
\bibitem{Gudnason:2006ug}
  S.~B.~Gudnason {\it et al}, \Journal{\PRD}{73}{115003}{2006}.
\bibitem{Gudnason:2006yj}
  S.~B.~Gudnason {\it et al}, \Journal{\PRD}{74}{095008}{2006}.

\bibitem{Norma}
N.S. Manko\v c Bor\v stnik, \Journal{\BWP}{11}{105}{2010}.
\bibitem{Norma2}
A. Bor\v stnik Bra\v ci\v
c, N.S. Manko\v c Bor\v stnik, \Journal{\PRD}{74}{073013}{2006}.
\bibitem{Norma3}
N.S. Manko\v c Bor\v stnik, \Journal{\MPLA}{10}{587}{1995}.
\bibitem{Norma4}
 N.S.
Manko\v c Bor\v stnik, \Journal{\IJTP}{40}{315}{2001}.
\bibitem{Norma5}
G. Bregar, M.
Breskvar, D. Lukman, N.S. Manko\v c Bor\v stnik,
\Journal{\NJP}{10}{093002}{2008}.

\bibitem{Levels1}
  M.~Y.~Khlopov, A.~G.~Mayorov and E.~Y.~Soldatov,
  \Journal{\BWP}{11}{73}{2010}.
\bibitem{KK2}
  M.~Y.~Khlopov and C.~Kouvaris, \Journal{\PRD}{78}{065040}{2008}.

\bibitem{unesco}
  M.~Y.~Khlopov, \Journal{\AIPCP}{1241}{388}{2010}.

\bibitem{iwara}
  M.~Y.~Khlopov, A.~G.~Mayorov and E.~Y.~Soldatov, \Journal{\IJMPD}{19}{1385}{2010}.

\bibitem{I2}
  M.~Y.~Khlopov,
  arXiv:0806.3581 [astro-ph].

\bibitem{Starkman}
  C.\,B. Dover {\it et al}, \Journal{\PRL}{42}{1117}{1979}.
  \bibitem{Wolfram}
  S. Wolfram, \Journal{\PLB}{82}{65}{1979}.
\bibitem{Starkman2}
G.\,D. Starkman  {\it et al}, \Journal{\PRD}{41}{3594}{1990}.
\bibitem{Javorsek}
 D.~Javorsek  {\it et al}, \Journal{\PRL}{87}{231804}{2001}.
\bibitem{Mitra}
S. Mitra, \Journal{\PRD}{70}{103517}{2004}.
\bibitem{Mack}
  G.~D.~Mack  {\it et al}, \Journal{\PRD}{76}{043523}{2007}.



\bibitem{McGuire:2001qj}
B.\,D. Wandelt et al.,
  arXiv:astro-ph/0006344.
\bibitem{McGuire2}
P.\,C. McGuire and P.\,J. Steinhardt,
  arXiv:astro-ph/0105567.
\bibitem{ZF}
 G. Zaharijas and G.\,R. Farrar, \Journal{\PRD}{72}{083502}{2005}.

\bibitem{Khlopov:2008rp}
  M.~Y.~Khlopov,
  arXiv:0801.0167 [astro-ph].
\bibitem{KhlopovPHE}
  M.~Y.~Khlopov,
  arXiv:0801.0169 [astro-ph].
\bibitem{DMDA}
 M.~Y.~Khlopov, A.~G.~Mayorov and E.~Y.~Soldatov, \Journal{\BWP}{11}{185}{2010}.

\bibitem{Bernabei:2003za}
  R.~Bernabei {\it et al.}, \Journal{\RNC}{26}{1}{2003}

\bibitem{Bernabei:2008yi}
  R.~Bernabei {\it et al.}  [DAMA Collaboration], \Journal{\EPHJ}{C56}{333}{2008}

\bibitem{Maltoni:1999ta}
M. Maltoni {\it et al.}, \Journal{\PLB}{476}{107}{2000}

\bibitem{Belotsky:2000ra}
K.M.Belotsky, M.Yu.Khlopov and K.I.Shibaev, \Journal{\GaCS}{6}{140}{2000}.

\bibitem{Belotsky:2005uj}
K.M.Belotsky {\it et al.}, \Journal{\GaC}{11}{16}{2005}.

\bibitem{Belotsky:2004st}
K.M.Belotsky {\it et al.}, \Journal{\PAN}{71}{147}{2008}.

\bibitem{Cahn}
R.~N.~Cahn and S.~L.~Glashow, \Journal{\SCI}{213}{607}{1981}.
\bibitem{Pospelov}
M.~Pospelov, \Journal{\PRL}{98}{231301}{2007}.
\bibitem{Kohri}
 K.~Kohri and
F.~Takayama,  \Journal{\PRD}{76}{063507}{2007}.

\bibitem{Mayorov}
K.M.Belotsky, A.G.Mayorov, M.Yu.Khlopov. Charged particles of dark
matter in cosmic rays. ISBN 978-5-7262-1280-7, Scientific Session
NRNU MEPhI-2010, V.4, P.127

\bibitem{Finkbeiner:2007kk}
  D.~P.~Finkbeiner and N.~Weiner, \Journal{\PRD}{76}{083519}{2007}

\bibitem{integral} B. J. Teegarden {\it et al}, \Journal{\APJ}{621}{296}{2005}

\bibitem{XQC}
D. McCammon  {\it et al}, \Journal{\NIMA}{370}{266}{1996};
\bibitem{XQC1}
D. McCammon {\it et al}, \Journal{\APJ}{576}{188}{2002}.


\bibitem{Belotsky:2006fa}
  K.~Belotsky {\it et al},
  arXiv:astro-ph/0606350.

\bibitem{Akerib:2005kh}
  D.~S.~Akerib {\it et al.}  [CDMS Collaboration], \Journal{\PRL}{96}{011302}{2006}.
\bibitem{Ahmed:2008eu}
  Z.~Ahmed {\it et al.}  [CDMS Collaboration], \Journal{\PRL}{102}{011301}{2009}.

\bibitem{cdms}
N.~Mirabolfathi {\it et al.}  [CDMS Collaboration], \Journal{\NIMA}{559}{417}{2006}.

\bibitem{xenon}
  E.~Aprile {\it et al.}  [XENON100 Collaboration], \Journal{\PRL}{105}{131302}{2010}.

\bibitem{cogent}
C. E. Aalseth {\it et al.}, \Journal{\PRL}{107}{141301}{2011}.

\bibitem{cresst}
G. Angloher {\it et al.},
arXiv:1109.0702 [astro-ph.CO].

\bibitem{Bled09}
M. Yu. Khlopov, A. G. Mayorov, E.Yu. Soldatov, \Journal{\BWP}{10}{79}{2009}.



\bibitem{DAMAlibra}
  R.~Bernabei {\it et al.}  [DAMA Collaboration], \Journal{\NIMA}{592}{297}{2008}

\bibitem{Edward}
  F.~Petriello and K.~M.~Zurek, \Journal{\JHEP}{0809}{047}{2008}.
 \bibitem{Foot}
R. Foot, \Journal{\PRD}{78}{043529}{2008}.
\bibitem{Feng1}
J. L. Feng, J. Kumar and
L. E. Strigari, \Journal{\PLB}{670}{37}{2008}.
\bibitem{Feng2}
J.~L.~Feng, J.~Kumar,
J.~Learned and L.~E.~Strigari, \Journal{\JCAP}{0901}{032}{2009}.
\bibitem{Drob}
  E.~M.~Drobyshevski, \Journal{\MPLA}{24}{177}{2009}.
\bibitem{Feldstein:2009tr}
  B.~Feldstein, A.~L.~Fitzpatrick and E.~Katz, \Journal{\JCAP}{1001}{020}{2010}.
\bibitem{Bai}
  Y.~Bai and P.~J.~Fox, \Journal{\JHEP}{0911}{052}{2009}.
\bibitem{Feldstein:2009np}
  B.~Feldstein {\it et al.}, \Journal{\JCAP}{1003}{029}{2010}.
\bibitem{Fitzpatrick:2010em}
  A.~L.~Fitzpatrick, D.~Hooper and K.~M.~Zurek, \Journal{\PRD}{81}{115005}{2010}.
\bibitem{Andreas:2010dz}
  S.~Andreas {\it et al.}, \Journal{\PRD}{82}{043522}{2010}.
\bibitem{Alves:2010dd}
  D.~S.~M.~Alves {\it et al.}, \Journal{\JHEP}{1006}{113}{2010}.
\bibitem{Barger:2010yn}
  V.~Barger, M.~McCaskey and G.~Shaughnessy, \Journal{\PRD}{82}{035019}{2010}.
\bibitem{Savage:2010tg}
  C.~Savage {\it et al.}, \Journal{\PRD}{83}{055002}{2011}.
\bibitem{Hooper:2010uy}
  D.~Hooper {\it et al.}, \Journal{\PRD}{82}{123509}{2010}.
\bibitem{Chang:2010pr}
  S.~Chang, R.~F.~Lang and N.~Weiner, \Journal{\PRL}{106}{011301}{2011}.
\bibitem{Chang:2010en}
  S.~Chang, N.~Weiner and I.~Yavin, \Journal{\PRD}{82}{125011}{2010}.
\bibitem{Barger:2010gv}
  V.~Barger, W.~Y.~Keung and D.~Marfatia, \Journal{\PLB}{696}{74}{2011}.
\bibitem{Fitzpatrick:2010br}
  A.~L.~Fitzpatrick and K.~M.~Zurek,
  arXiv:1007.5325 [hep-ph].
\bibitem{Banks:2010eh}
  T.~Banks, J.~F.~Fortin and S.~Thomas,
  arXiv:1007.5515 [hep-ph].
\bibitem{Feldstein:2010su}
  B.~Feldstein, P.~W.~Graham and S.~Rajendran, \Journal{\PRD}{82}{075019}{2010}.

\bibitem{Gelmini}
    G.~B.~Gelmini, \Journal{\IJMPA}{23}{4273}{2008}.
\bibitem{Aprile:2009zzd}
  E.~Aprile, S.~Profumo, \Journal{\NJP}{11}{105002}{2009}.
\bibitem{Feng:2010gw}
  J.~L.~Feng, \Journal{\ARAA}{48}{495}{2010}.
\bibitem{Dai}
De-Chang Dai, K. Freese, D. Stojkovic, \Journal{\JCAP}{0906}{023}{2009}.

\bibitem{Jia}
Jia-Ming Zheng et al., \Journal{\NPB}{854}{350}{2012}.


\end{thebibliography}
\end{document}